\documentclass{aadebug}
\corr{0309027}{13}

\newcommand{\acsc}[1]{
	\ifthenelse{
		\equal{\version}{acsc}
	}{
		{#1}
	}{
		{}
	}
}
\newcommand{\iclp}[1]{
	\ifthenelse{
		\equal{\version}{iclp}
	}{
		{#1}
	}{
		{}
	}
}

\newcommand{\longex}[1]{
	\ifthenelse{
		\equal{\whichexample}{long}
	}{
		{#1}
	}{
		{}
	}
}
\newcommand{\shortex}[1]{
	\ifthenelse{
		\equal{\whichexample}{short}
	}{
		{#1}
	}{
		{}
	}
}

\begin{document}

\title{Idempotent I/O for safe time travel}

\author{
Zoltan Somogyi\addressnum{1}
}

\address{1}{zs\@@cs.mu.OZ.AU \\
Department of Computer Science and Software Engineering, \\
University of Melbourne, Victoria, 3010, Australia \\
Phone: +61 3 8344 1300, Fax: +61 3 9348 1184}

\pdfinfo{
/Title (Idempotent I/O for safe time travel)
/Author (Zoltan Somogyi)
}

\begin{abstract}
Debuggers for logic programming languages have traditionally had
a capability most other debuggers did not:
the ability to jump back to a previous state of the program,
effectively travelling back in time in the history of the computation.
This ``retry'' capability is very useful,
allowing programmers to examine in detail
a part of the computation that they previously stepped over.
Unfortunately, it also creates a problem:
while the debugger may be able to restore the previous values of variables,
it cannot restore the part of the program's state
that is affected by I/O operations.
If the part of the computation being jumped back over performs I/O,
then the program will perform these I/O operations twice,
which will result in unwanted effects ranging
from the benign (e.g. output appearing twice)
to the fatal (e.g. trying to close an already closed file).

We present a simple mechanism for ensuring that
every I/O action called for by the program is executed at most once,
even if the programmer asks the debugger to travel back in time
from after the action to before the action.
The overhead of this mechanism is low enough and can be controlled well enough
to make it practical to use it to debug
computations that do significant amounts of I/O.
\end{abstract}


\section{Introduction}
\label{sect:intro}

Programmers often have the following experience when debugging a program:
\begin{enumerate}
\item
They check the values of a procedure's input arguments,
and find them to be all OK.
\item
They step over the execution of the procedure in the debugger,
regaining control when the procedure returns to its caller.
\item
They check the values of the procedure's outputs,
and find some of them to be in error.
\end{enumerate}

At this point, they know there is an error
somewhere inside the call-tree of the procedure,
but they don't yet know precisely where.
The natural action to find out would be to reexecute the call
and check the values of variables at a selection of program points
\emph{before} the call returns.

Traditional debuggers such as gdb
do not help the programmer to perform the natural action.
They can execute the program only in the usual direction: forwards.
Once an assignment statement has been executed,
there is generally no way to recover
the previous value of the assigned-to variable.
Restoring the computation to the state it had before the call
is therefore not possible.
The only action the programmer can take
is to start the program again from scratch,
stop its execution at the problematic call,
and examine the execution of the call in more detail.
This has several problems.
\begin{itemize}
\item
If the program modifies data it uses as input,
as most programs that manipulate databases do,
then the programmer must restore the database to its initial state
before he or she can reexecute the program.
\item
The part of the program before the problematic call
may take considerable time to reexecute.
\item
Identifying the problematic call may be a tricky task in its own right,
because the programmer may have arrived at that call
after a long sequence of operations (continue to breakpoint, skip, next etc)
that he or she may be unwilling or unable to repeat.
\item
That reexecution may require the programmer to type in input
they have typed in before when that part of the program was being executed
for the first time.
\item
Any accidents such as typos that cause the program to be given
different input on reexecution than it had on first execution
may (and probably will) prevent the debugger from reestablishing
the state of the computation at the time of the call.
The only cure is to restart execution one more time.
\item
If some of the input to the program comes from sources
that are outside the programmer's control,
such as network connections or the precise timing of input operations,
then reestablishing the state of the computation at the time of the call
may never be possible.
\end{itemize}
A mechanism that would allow the debugger
to reset the computation to the state it had at the time of the call,
effectively allowing the programmer to jump backwards
in the program's timeline, would avoid these problems.

In section 2, we review existing work
on providing debuggers with the ability to travel back in time.
In section 3, we give examples of time travel across I/O actions
that existing systems cannot handle correctly.
In section 4, we describe a mechanism that enables time travel to work
even when jumping over I/O actions.
In section 5, we present some related work.

\section{Existing implementations of time travel}
\label{sect:timetravel}

Debuggers for imperative languages could, in theory,
log the old value of the assigned-to variable before each assignment,
and then play the log backwards to restore variables
to the values they had at any previous point in time.
However, since today's CPUs can execute a billion assignments in a second,
the log will quickly outgrow all reasonably sized storage devices,
and noone has yet found a way to compress it to a feasible size.

Debuggers for declarative languages do not have any problem with log sizes,
because their variables are single-assignment:
once a variable has been given a value, that value cannot be changed.
All variables that were bound when the call was made
will therefore still have the same value when the call returns.
(Some optimizations can reuse the memory storing a variable's value
to store something else after that variable has become dead,
but such optimizations can be switched off.)
To restore \emph{all} variables to the state they had at the time of the call
they do have to find out which variables were unbound at the time of the call,
and reset them to that state if the call bound them.

Fortunately, this is not hard.
Even in the absence of considerations of debugger support,
most logic programming languages must have a mechanism
to solve a very closely related problem:
resetting variables to unbound on backtracking.\footnote{
The exceptions are languages that implement don't know nondeterminism
through a mechanism other than backtracking, e.g. OR-parallelism,
or don't implement it at all.}
In Prolog, the mechanism is the trail,
a kind of log of the addresses of variables bound on forward execution
\cite{wamtute}.
In Mercury, it is the strong mode system that allows the implementation
to avoid reading the ``value'' of a variable before it becomes bound,
making ``reset to unbound'' a null operation and avoiding the need for a trail
\cite{jlp}.
Functional languages are in effect strongly moded,
with function arguments being inputs and return values being outputs.

Both these mechanisms can be easily adapted for use by the debugger.
The adaptation is simplest if the debugger allows backward jumps in time
only if the destination represents the call of a currently active procedure.
Allowing arbitrary destinations would require more overhead
e.g. for taking snapshots of the trail pointer at more program points,
and any mechanism that would allow the programmer to specify
an arbitrary destination would complicate the debugger's user interface.
Traditionally, jumping back in time to the start of a call
is called a \emph{retry} of that call.
Specifying the active ancestor to retry
requires simply picking that ancestor off a list of ancestors,
and most of the time that is exactly what the programmer wants to do anyway.
At other times, programmers can retry the closest ancestor
whose start time is before the time they want to jump to
and execute forward from there.
This leaves them vulnerable to the problems involved
in restarting the program from scratch, but to a smaller extent.

Retry operations in debuggers can be compromised
by non-declarative constructs in otherwise declarative languages.
For example, one can think of Prolog's assert/retract operations
as destructive assignments to a variable or set of variables
representing the clause database.
Since they do not keep logs of updates to the clause database,
Prolog debuggers cannot restore its previous states.
Their implementations of retry operations are therefore flawed:
even though they can restore the the variables to their required states,
they cannot guarantee that execution will follow the same path after a retry
as when the retried program fragment was first executed.
We call a retry operation \emph{safe} only if it doesn't allow
execution to ``get lost'' in such a manner. Only safe retries guarantee that
programmers will be able to examine the computations they want to debug.

Unfortunately, the mechanisms we have discussed in this section
do not guarantee safety even for purely declarative languages,
since even programs written in such languages
must perform I/O if they are to be useful,
and these mechanisms do not address the problem of retries across I/O actions.

\section{The problem of retry across I/O actions}
\label{sect:procprob}

If the computation executes I/O operations
between the start of the call being retried
and the point where the programmer asks the debugger to perform the retry,
then the retry will cause those I/O operations to be executed
for a second time (and maybe a third, fourth etc time).

Sometimes the effect of this reexecution is trivial;
sometimes it is grave.
We give four examples illustrating the range.
The examples are in Prolog syntax, with some pseudocode.

\begin{verbatim}
write_solution(ProblemDescription) :-
    <compute Solution from ProblemDescription>
    write(Solution).
\end{verbatim}

The effect of reexecuting \texttt{write\_solution}
is that some output is duplicated.
If output is going to the programmer's screen, it can simply be ignored.
If output is going to a file, it may also be ignored,
but duplicated output segments will prevent automated comparisons
between the program's expected and actual outputs
from producing meaningful results.

\begin{verbatim}
read_problem(Solution) :-
    read(ProblemDescription),
    <compute Solution from ProblemDescription>.
\end{verbatim}

The effect of reexecuting \texttt{read\_problem}
is that the program requires \texttt{ProblemDescription} to be input twice.
If the input is coming from the keyboard,
this can be anywhere from annoying (if only a few keystrokes are required)
to impossible
(if a thousand keystrokes are required, a typo is almost guaranteed).
If the input is coming from a file,
the programmer must edit the file on the fly,
and remember to restore its contents afterward.
If the input is coming from another computer through a network connection,
providing the same input again may be impossible to arrange.

\begin{verbatim}
get_stream(Stream) :-
    write("please type filename: "),
    read(Filename),
    open(Filename, read, Stream).
\end{verbatim}

Besides reexecuting the read and write operations,
reexecuting \texttt{get\_stream} opens the file twice.
This represent a resource leak,
the resource being the data structures used by the OS to represent open files
(e.g. file descriptors in Unix).
These resources are typically finite in number,
so allocating one and then discarding it (as the retry operation does here)
can cause the program to eventually run out,
causing the failure of a later attempt to allocate a resource of the same type.
That attempt maybe in a part of the program
totally unrelated to predicate \texttt{get\_stream}.

\begin{verbatim}
read_next_item(Stream, MaybeItem) :-
    read(Stream, Item),
    ( Item = end_of_file ->
        MaybeItem = no,
        close(Stream)
    ;
        MaybeItem = yes(Item)
    ).
\end{verbatim}

If a call to \texttt{read\_next\_item} gets an end-of-file indication,
then reexecuting that call will try to close \texttt{Stream} twice.
Since operating systems allow streams to be closed only once,
the second attempt will fail.
If the implementation is strict about unexpected errors,
that failure may cause the program to be aborted then and there.

There are two reasons why I/O operations are problematic for retries.

The first reason is that I/O operations
are inherently destructive updates of the state of the world.
One can give I/O a declarative semantics via several techniques
(e.g. monads in Haskell, uniqueness types in Clean,
and unique modes in Mercury),
but all these effectively pretend that an I/O operation
represents a relationship between two single-assignment variables
representing the state of the world before and after the operation.
The actual implementation isn't single-assignment,
and for the debugger, it is the implementation that matters;
basing the language semantics on prohibiting access to past states of the world
doesn't help when you want to recreate those past states.
We call this the destructive I/O problem.

The second reason is that I/O operations affect resources
that are at least potentially also accessible to outside influences.
We can divide the state of the program into two parts:
the private state,
over which the program's compiler and the debugger together have total control,
and the public state,
which they may be able to affect but which they do not control.
This classification is related to but conceptually separate from
the one that separates the internal state,
which the program can affect directly,
from the external state,
which the program can affect only via the operating system.
The contents of memory and registers are part of the internal state
while the contents of the files and communication channels
accessed by the program
and of the kernel tables related to the program
are part of its external state.
All the internal state is normally private
(we will cover one exception in section 4),
and the external state at least potentially public.
Even if the debugger were able to take
a snapshot of the external state at a call,
(which is difficult enough if the program accesses a network),
and was able to restore that state when retrying that call
(which may require permissions that the program may not have),
there is no guarantee that some other entity (another process, the kernel)
won't perturb that state before the program being debugged accesses it.
We call this the volatile resources problem.

One could try to address the destructive I/O problem using the technique
we mentioned in section~2 as a potential solution for destructive assignments:
logging each I/O operation and undoing its effects
when a retry operation jumps over the operation.
This should be feasible because a typical program
executes far fewer I/O operations than assignments.
To make it possible, I/O operations must of course be reversible.
Some are: output to a screen can be painted over,
characters consumed from an input queue can be put back into the queue,
an opened file can be closed, a closed file can be opened again.
Some that are not
(e.g. sending a command to an ATM to dispense money
or sending your resignation to your boss by email),
while sometimes necessary, can hopefully be avoided during debugging,
but others (e.g. reading from a pipe connected to an external server)
may not be avoidable at reasonable cost.

One could try to avoid the volatile resources problem
by making all the resources needed by the process being debugged
private to the process.
One can do this by giving testers their own copies
of the files, databases, server processes etc needed by the program;
if all these resources are on a machine controlled exclusively by the tester,
then the tester may be able to ensure that the only thing accessing these
resources while the program is being debugged is the program itself.
Most books on software engineering encourage this in any case,
as a means of an effort to eliminate Heisenbugs,
bugs that cannot be reproduced reliably.
Unfortunately, making public external resources private
doesn't also make them internal.
As long as the resources manipulated by I/O operations
can be manipulated \emph{only} by I/O operations,
restoring those resources to a previous state
requires examining all types of I/O operations,
and designing and coding an algorithm that
undoes all the effects of that type of I/O operation
that are visible to the rest of the program.\footnote{
If the program depends e.g. on the success or failure of a file open operation
but not on the precise value of the returned file descriptor or file handle,
then it is OK for the open to succeed
with a different file descriptor or file handle after a retry.}
In the best case, this requires great engineering effort.
In the worst case, it may be impossible;
for example, there is no way to get Unix to reset the kernel data structures
whose contents are returned by the \texttt{getrusage()} system call,
which reports on the resource consumption of the process so far.

\section{Making I/O actions idempotent}
\label{sect:idempotent}

The key observation of this paper is that
while restoring external resources to a previous state may be difficult,
we do not actually need to do it;
it is sufficient to make the program behave as if we had done it.
The insight required to exploit this observation
is that the program interacts with those external resources
through a limited number of I/O operation types,
and the program being debugged doesn't care what these operations do
as long as they return the right results.

The obvious question is: how do you know whether
the results of the operations executed by the program are right?
One obvious answer is: if the operations executed by the program
return results that coincide with the results that could be returned
by an execution of the program that does not perform any retries,
one may as well accept those results as correct.\footnote{
Note that we say ``\emph{an} execution of the program''
not ``\emph{the} execution of the program''.
Small timing variations can affect the order of interleaving
of the operations on shared resources
executed by the process being debugged and other processes,
and debuggers always have some timing overhead,
so no debugger can do better.
Of course, one can reduce the impact of this effect
by reducing the number of resources that
the program being debugged shares with other processes.}

We can guarantee that all I/O operations return results
that are correct by this measure, even in the presence of retries,
by making all I/O operations \emph{idempotent}.
This means that
\begin{itemize}
\item
when the program executes an I/O operation for the first time,
we execute the actions called for by the operation and record the results;
\item
when the program executes an I/O operation for the second, third etc time,
after a retry has warped time from after the operation to before it,
we just return the results we recorded the first time
without actually performing any I/O.
\end{itemize}

We have designed a simple mechanism to enforce idempotence of I/O operations
and integrated it into the Mercury implementation.
While we present the mechanism in the context of Mercury \cite{mercury_refman},
adapting the mechanism to other languages should be straightforward.

In Mercury, a predicate can perform I/O only if
it is guaranteed to succeed exactly once
and it has a pair of arguments
representing the state of the world outside the program
before and after the execution of the predicate,
with the argument representing the initial state of the world
having mode \texttt{di} (short for ``destructive input'')
and the argument representing the final state
having mode \texttt{uo} (short for ``unique output'').
The determinism requirement ensures that
Mercury programs never try to backtrack past an I/O operation,
while the mode requirement ensures that
programs can refer only to the current state of the world,
and cannot refer to to past states.

The Mercury standard library ultimately implements all I/O operations
using the foreign language interface.
Consider the predicate \texttt{read\_char},
whose task is to attempt to read a character from the specified stream
and to return a result specifying either
the character read (if the attempt is successful)
or an end-of-file or error indication (if the attempt failed).
It is written in Mercury, but it calls
a lower level predicate to do the actual I/O;
the Mercury code just interprets the integer code returned by 
this predicate, \texttt{read\_char\_code}.
To make I/O operations idempotent,
it is sufficient to impose idempotence on I/O primitives
(i.e. on predicates that perform I/O and are implemented in foreign code).

\texttt{read\_char\_code} is a typical I/O primitive,
with one input and one output besides the usual pair of I/O states;
the input specifies the stream to read from
and the output encodes the result of attempting to read one character
(ASCII values standing for the corresponding character,
-1 representing end of file, and all other values representing errors):

\begin{verbatim}
:- pred read_char_code(stream::in, int::out, io__state::di, io__state::uo)
    is det.
\end{verbatim}

For the C backend of Mercury, the implementation of this predicate is in C:

\begin{verbatim}
:- pragma foreign_proc("C",
    read_char_code(Stream::in, CharCode::out, S0::di, S::uo),
    [promise_pure],
"
    CharCode = getc(Stream);
    S = S0;
").
\end{verbatim}

The C code just calls \texttt{getc}.
The I/O state arguments are dummies,
standing in for and representing the external state of the process
but containing no meaningful values themselves;
the only reason why we assign to \texttt{S} is to avoid compiler warnings.
The \texttt{promise\_pure} annotation promises the compiler that
even though the implementation of \texttt{read\_char\_code} is imperative code,
that imperative code implements a declarative interface,
and that calls to \texttt{read\_char\_code}
can be reordered with respect to other goals
to the full extent allowed by data dependencies.
Data dependencies involving I/O states ensure that such reordering
will not alter the order in which I/O operations are carried out.

The Mercury compiler considers foreign code fragments like the one above
to be a type of primitive goal, just like calls and unifications.
The internal representation of the definition of \texttt{read\_char\_code}
is therefore something like this, although in actuality
the compiler records more information about foreign code goals:

\begin{verbatim}
read_char_code(Stream, CharCode, S0, S) :-
    <foreign_code, "C", [Stream, CharCode, S0, S],
        "CharCode = getc(Stream); S = S0;">.
\end{verbatim}

We say that each call to an I/O primitive is an \emph{I/O action}.
As part of making I/O actions idempotent,
we associate a sequence number with every one.
We do this by transforming the bodies of I/O primitive predicates,
and making the first action of the transformed code be
the allocation of a I/O action number for the current call.
The allocation is done by code in a builtin foreign language predicate
that simply increments a global counter variable
that holds the sequence number of the next I/O action to be executed.
\footnote{This assumes that the program is single-threaded.
The Mercury debugger does not (yet) support debugging
of multi-threaded programs.}

In the absence of retries, every call to an I/O primitive
will be allocated an I/O action number that hasn't been seen before
during this execution of the program.
However, this is not true in the presence of retries.
When generating debuggable executables, the Mercury compiler
extends all stack frames with an extra slot,
and copies the current value of the global I/O action counter variable
into this slot when the stack frame is first created at a predicate call.
The compiler also records the location of this stack slot
in the runtime type information (RTTI) it generates for the debugger.
This way, when the debugger performs a retry,
it can (and does) reset the global I/O action counter variable
to the value it had on entry to the call being retried.
If the retry jumps backward over $N$ I/O actions,
this will decrement the global counter by $N$,
which means that the next $N$ calls to I/O primitives
will be allocated I/O action numbers that have been seen before.
If we can ensure that execution will take the same path after the retry
that it had taken after the original invocation of the call being retried
(and we do ensure this, with some cooperation from the programmer),
then a given I/O action number will identify the same action
after the retry as it did before the retry.

The idempotence transformation uses a data structure, the I/O action table,
that maps the I/O action number of every action executed by the program so far
to the values of the output variables of the I/O primitive predicate
involved in that action.
The data structure we have chosen is a simple array
which is reallocated with a doubled size whenever it needs to expand,
since this is faster than and at least as space efficient
as structures based on trees.
The entries in the array corresponding to I/O actions that have occurred
will point to a block of memory, with each word in that block
containing the value of an output argument of the corresponding action;
the other entries in the array will contain a null pointer.

One way to implement idempotence is via a source-to-source transformation
that converts the definition of \texttt{read\_char\_code} shown above
into the following;
the \texttt{impure} and \texttt{semipure} markers are explained below.

\begin{verbatim}
read_char_code(Stream, CharCode, S0, S) :-
    impure allocate_io_action_number(IoActionNumber),
    ( semipure io_has_occurred(IoActionNumber, ResultBlock) ->
        semipure restore_answer(ResultBlock, 0, CharCode),
        semipure restore_answer(ResultBlock, 1, S)
    ;
        <foreign_code, "C", [Stream, CharCode, S0, S],
            "CharCode = getc(Stream); S = S0;">,
        impure create_answer_block(IoActionNumber, 2, ResultBlock),
        impure save_answer(ResultBlock, 0, CharCode),
        impure save_answer(ResultBlock, 1, S)
    ).
\end{verbatim}

This code allocates an I/O action number for this action,
and checks whether this is the first execution of this action.
If it is, we execute the original body of the predicate,
and then record the values of its output arguments
in a result block pointed to from this action's entry in the I/O action table.
If the action has been executed before,
we do not execute the original body;
instead, we just look up the values of the output arguments we need to return
in the result block allocated by the first execution.
The transformation algorithm is straightforward:

\begin{itemize}
\item
The first goal in the else branch is the original body.
\item
The first two lines of the transformed code are fixed,
as is the call to \texttt{create\_answer\_block}
except for the second argument.
\item
The second argument \texttt{create\_answer\_block} and
the blocks of calls to \texttt{restore\_answer} and to \texttt{save\_answer}
are derived in the obvious fashion
from the list of the predicate's output arguments.
\end{itemize}

Note that the allocation of an I/O action number,
the allocation of a new entry in the I/O action table
and the filling in of that entry all modify global data structures,
while the test for the I/O action having occurred
and the code for restoring previously computed answers
read those global data structures.
This is why the goals in the first class are marked as impure,
while the goals in the second class are marked as semipure \cite{purity}.
The compiler is allowed to swap the order of two semipure goals,
but it cannot swap two impure goals or one impure and one semipure goal.
Those markings therefore ensure that the updates and tests of global variables
executed by the foreign code implementations of the predicates
called by the code inserted into I/O primitives take place in the proper order,
without restricting the Mercury compiler's freedom to reorder code
any more than necessary.
Note that the transformed body of the predicate as a whole is a pure goal
even though contains impure and semipure components.
A goal is pure if its outputs depend only on its inputs,
which in this case include an I/O state
representing the external state of the program;
the side effects that the transformed procedure body executes and relies on
are not detectable by the rest of the program.

The transformation actually performed by the Mercury compiler
is a bit more complicated than the one above, for several reasons.

The first is that even before we started working on idempotent I/O,
the Mercury standard library had predicates
for manipulating answer blocks and sets of key/value pairs;
they were used to implement tabling
(also known as memoization or automatic caching).
Instead of implementing the auxiliary predicates
called by the idempotency transformation from scratch,
we used and adapted the existing implementations,
some of which have a slightly different interface.
Due to this reuse of the tabling infrastructure,
and the fact that its main data structure is a table,
we refer to our idempotency transformation as \emph{I/O tabling}.

The second difference is that we store
not just the output arguments of I/O actions,
but also their input arguments and the identity of the predicate involved.
This extra information is not necessary to make retry safe,
but the Mercury declarative debugger needs it to handle predicates that do I/O.
Normally, declarative debuggers \cite{Shapiro1983,Lloyd1987b}
ask programmers questions about whether the output argument values
computed by a call to a predicate represent
a correct set of outputs given the values of the input arguments.
However, this doesn't work on predicates that do I/O.
For predicates that read input, the correctness of the output arguments
depends not only on the input arguments but also on what the call read;
for predicates that write output, the correctness of the predicate
depends not only on the output arguments but also on what the call wrote.
The information we record allows the declarative debugger
to record for each call the value of the I/O action counter
on entry to the call and on exit from the call.
Together with our extended table, these implicitly represent
the list of I/O actions executed by the call.
When asking about the correctness of a call,
the Mercury declarative debugger can materialize this list
for display to the programmer.\footnote{
When the list is too long to display,
we use the technique we also use
for argument terms that are too large to display:
we display a part and let the programmer browse the rest.}

The third difference is that when an argument is of a builtin type,
we use specialized, monomorphic versions of
\texttt{save\_answer} and \texttt{restore\_answer} for it
so as not to incur the overhead of polymorphic calls.

The fourth difference is that since I/O state arguments are just dummies,
we do not save and restore them.
This reduces overheads in both time and space.

The fifth difference is that we can table more than just I/O operations.
This is necessary because the foreign language interface of Mercury
can of course be used for purposes other than implementing I/O.
Some predicates implemented in foreign code represent pure computations;
examples include predicates that just call
a mathematical functions such as \texttt{sin()} or \texttt{log()}.
Some predicates implemented in foreign code
represent impure or semipure computations;
examples include predicates that execute database transactions or queries.
Foreign language predicates in the first category are declarative,
and behave the same way with respect to retries
as predicates implemented in Mercury code.
Foreign language predicates in the second category are not declarative,
and their actions are outside the control of the Mercury compiler and debugger.
In effect, they access a part of the program's state
that is internal and yet (by our definition) not private.
They pose the same problems on retries as I/O primitives,
and we can handle them the safe way, using the idempotency transformation.
At the moment, we require programmers to mark
impure and semipure foreign language predicates
with an annotation that tells the compiler
to perform the idempotency transformation on them.
If the programmer does this,
then I/O tabling can guarantee the safety of all retries,
provided of course it is turned on.

That proviso exists because the last difference is that
our transformation makes I/O tabling optional,
and allows programmers to choose between three alternatives at runtime:

\begin{itemize}
\item
I/O tabling is not turned on at all.
This corresponds to the situation before I/O tabling was implemented,
with the exception that all I/O primitives perform an extra test
(which always fails) to check whether I/O tabling is enabled.
This is the most efficient option,
but it provides no safety for retries across I/O actions.
\item
I/O tabling is turned on throughout the program's execution.
This provides complete safety for retries across I/O actions,
but it is the least efficient option in terms both time and space.
The direct memory overhead of the I/O action table
is roughly proportional to the number of I/O actions executed by the program;
the proportion is not exact because of our doubling reallocation policy
and because the answer blocks of different I/O primitives have different sizes.
The appearance of a term in an answer block
prevents the memory of that term from being garbage collected;
we believe this overhead also tends to be roughly proportional
to the number of I/O actions.
\item
The programmer turns on I/O tabling some time after the program starts
and turns off I/O tabling some time before the program exits,
dividing the program's timeline into three regions:
before, during and after I/O tabling.
The overhead of this option is proportional
to the number of I/O actions executed by the program in the ``during'' region.
Retries are safe only if both endpoints are within the ``during'' region.
If a retry is unsafe, the debugger warns the programmer
and allows the retry to be aborted.
\end{itemize}

One of our standard benchmarks is the Mercury compiler,
which is itself written in Mercury, compiling six of its largest modules,
which total over 34,000 lines and 1.2~Mb.
On a PC with a 2.4~GHz Pentium 4 CPU and 512~Mb of memory,
this takes about 190 seconds when running the compiler
inside the Mercury debugger with I/O tabling disabled.
Turning on I/O tabling for the whole run
causes this time to rise only to about 210 seconds, an overhead of only 10\%,
even though the benchmark executes over 12 million I/O actions.
The 12+ million entry I/O table roughly triples
the compiler's memory requirements,
taking them from about 127~Mb to about 389~Mb.
(There is only one relevant memory size
because the Mercury debugger is part of the same address space
as the program being debugged \cite{mercdebug}.)

We have provided the ability to turn on I/O tabling
only for part of the program's execution
to cater for programs that are significantly
longer-running and/or more I/O intensive than the compiler.
However, we do not expect this ability to be required all that often,
because we expect bugs that require that big a test case to show themselves
to be reasonably rare.
We could reduce memory overheads significantly 
(for this benchmark, to about 291~Mb)
by storing only output arguments in the I/O table,
but we judge support for declarative debugging to be more important
than increasing the size of the test case required
to cause the program being debugged to start thrashing.

\section{Related work}
\label{sect:related}

There have been many implementations of time travel in debuggers
in the last thirty years or so.
The following discussion is not meant to be exhaustive.
Interested readers will find a good historical survey in \cite{tolmachdebug};
more recent work, e.g. for Java \cite{bdbj}, can be found on the Web.

Most implementations of time travel
have been limited to restoring internal state only.
The central concern when restoring internal state
is avoiding excessive growth of the undo log.
Debuggers aimed at educational settings, e.g. Leonardo \cite{leonardo},
may be able to avoid facing the issue:
since the programs of concern tend to have short executions,
the log tends to be small too.
Debuggers aimed for more general use have no such option.
One alternative their designers can take
is to simply discard old log entries, at the cost of
allowing time travel back only by a limited number of instructions,
as in Borland's Turbo Debugger \cite{turbodebug}.
Another is to replace old log entries (or maybe all of them)
with one or more checkpoints (snapshots) of the program state;
to travel back to a region of the execution not covered by log entries,
the debugger restores the last checkpoint before the desired time point
and executes the program forward from there.
The choice of the number of checkpoints to maintain
is a classic time/space tradeoff:
increasing the number of checkpoints
reduces the average amount of reexecution required.

This tradeoff is more favourable in impure declarative languages such as ML,
because only the impure part of the program needs logging or checkpointing
\cite{tolmachdebug}.
The pure part of the program works with immutable data structures,
which by definition are written to only when created.
Preparing to travel back in time
across the creation of an immutable data structure
does not impose extra overhead
either because the implementation knows which variables are bound when
and can thus undo their binding without log entries (as in functional languages)
or because it already incurs the cost of a log-like data structure
to support backtracking (such as the trail in Prolog).
The SML/NJ debugger \cite{tolmachdebug} shows that
the cost of checkpointing the mutable state can be acceptable,
at least for programs that are mostly declarative.
Unfortunately, Prolog debuggers based on the box model \cite{byrd80}
typically do not ever undo asserts/retracts, I/O and other side effects
when executing retries, nor do they try to avoid reexecuting them.
Prolog programmers have therefore had to live
with retries over impure code being unsafe.

The program reexecution required by checkpointing system
is safe only if the implementation avoids redundant executions
of actions that modify external and/or public state.
Only a few systems have attempted to ensure this.
One, Leonardo \cite{leonardo}, actually erases
the visible effects of I/O actions when jumping back across their executions.
This is possible because Leonardo's programs run on a private virtual machine
fully controlled by Leonardo's implementors;
while the state is external to the program being debugged,
it is internal to the program implementing the virtual machine,
and can thus be private.
Another system, the SML/NJ debugger, makes I/O operations idempotent
by using special versions of the relevant library functions.
That makes this the closest systems to ours.
The main differences we know about are that
the SML/NJ system doesn't support declarative debugging,
that it does support debugging of multithreaded programs,
and that it uses two different mechanisms
to handle impure code (code that uses mutable variables) and I/O.
Our use of a single mechanism to handle impure code
(in our case, calls to predicates implemented in other languages) and I/O
may be slightly less efficient
(specialized log records, table entries etc can be smaller than general ones),
but it makes the implementation simpler.
Unfortunately, the papers describing the SML/NJ debugger
\cite{tolmachdebug,tolmachthesis} don't say
how I/O functions are made idempotent
(what the transformation is, whether it is automatic,
and whether it can be applied to foreign language calls
in an implementation that supports such calls),
so we cannot compare it with our own transformation in more detail.

Tools such as Xrunner \cite{xrunner}\footnote{
The company selling Xrunner, Mercury Interactive,
and the Mercury programming language are not related in any way
beyond sharing a name.} can capture and replay
classes of I/O actions such as those related to GUI events.
However, their intended use is test automation;
we are not aware of any such tool being integrated into a debugger.

\section{Conclusion}
\label{sect:conc}

Our experience and that of many others' shows that
the ability to travel backward in time
can significantly improve a programmer's productivity when chasing a bug.
However, if the path of execution after the time jump
doesn't match its initial path,
this productivity boost may disappear or even turn negative,
as the programmer may then be led to explore and attempt to debug
a program state that would not have arisen using only forward execution.
The utility of a retry operation therefore depends on its safety.

Jumping backward in time across an action executed by the program
is safe only if the debugger can undo the effect of the action,
or can simulate having done so.
Some types of actions
including I/O operations and executions of foreign language code,
cannot be feasibly undone, so their undoing must be simulated:
the language implementation must ensure that
when forward execution resumes after a retry,
it will be behave \emph{as if} those actions were undone.
The idempotency transformation is a simple and general mechanism
for implementing this simulation.

We have implemented the idempotency transformation
for use in the Mercury debugger,
and shown that its overhead is low enough
to make it practical to use it to debug
computations that do significant amounts of I/O.
We have also used it to allow the Mercury declarative debugger 
to handle programs that do I/O.
However, the technique is not unique to Mercury,
and should be adaptable to both imperative and declarative languages.

We would like to thank Fergus Henderson
for the idea of reusing Mercury's tabling infrastructure
in the implementation of the idempotency transformation.

We would like to thank the Australian Research Council and Microsoft
for their support.

\bibliography{idempotent}
\end{document}